\documentclass[aps,prd,preprint,superscriptaddress,tightenlines,%
nofootinbib]{revtex4}
\newcommand{\PRE}[1]{{#1}}   

\usepackage{bm}
\usepackage{epsfig}
\usepackage{latexsym}

\newcommand{\postscript}[2]{\setlength{\epsfxsize}{#2\hsize}
   \centerline{\epsfbox{#1}}}

\def\k{\kappa}                    

\def\m{\mu}
\def\n{\nu}
  
\def\6{\partial}

\def\ed{\end{document}}

\newcommand{\be}{\begin{equation}}
\newcommand{\ee}{\end{equation}}
\newcommand{\beq}{\begin{equation}}
\newcommand{\eeq}{\end{equation}}
\newcommand{\bea}{\begin{eqnarray}}
\newcommand{\eea}{\end{eqnarray}}

\begin{document}

\preprint{
\hfil
\begin{minipage}[t]{3in}
\begin{flushright}
\vspace*{.4in}
\end{flushright}
\end{minipage}
}

\title{
\PRE{\vspace*{1.5in}} Cosmology from String Theory
\PRE{\vspace*{0.3in}} }

\author{Luis Anchordoqui}
\affiliation{Department of Physics,\\
University of Wisconsin-Milwaukee,
 Milwaukee, WI 53201
\PRE{\vspace*{.1in}}
}

\author{Haim Goldberg}
\affiliation{Department of Physics,\\
Northeastern University, Boston, MA 02115
\PRE{\vspace*{.1in}}
}

\author{Satoshi Nawata}
\affiliation{Department of Physics,\\
University of Wisconsin-Milwaukee,
 Milwaukee, WI 53201
\PRE{\vspace*{.1in}}
}

\author{Carlos Nu\~nez}
\affiliation{
Department of Physics,\\
University of Swansea, Singleton Park, Swansea SA2 8PP, UK
\PRE{\vspace*{.5in}} }

\date{April 2007}
\PRE{\vspace*{.5in}}
\begin{abstract}

\noindent We explore the cosmological content of Salam-Sezgin six
  dimensional supergravity, and find a solution to the field equations
  in qualitative agreement with observation of distant supernovae,
  primordial nucleosynthesis abundances, and recent measurements of
  the cosmic microwave background. The carrier of the acceleration in
  the present de Sitter epoch is a quintessence field slowly rolling
  down its exponential potential. Intrinsic to this model is a
  second modulus which is automatically stabilized and acts as a source
  of cold dark matter, with a mass proportional to an exponential
  function of the quintessence field (hence realizing VAMP models
  within a String context). However, any attempt to saturate the
  present cold dark matter component in this manner leads to
  unacceptable deviations from cosmological data -- a numerical study
  reveals that this source can account for up to about 7\% of the
  total cold dark matter budget. We also show that (1) the model will
  support a de Sitter energy in agreement with observation at the
  expense of a miniscule breaking of supersymmetry in the compact
  space; (2) variations in the fine structure constant are controlled by
  the stabilized modulus and are negligible; (3) ``fifth''
  forces are carried by the stabilized modulus and are short range;
  (4) the long time behavior of the model in four dimensions
  is that of a Robertson-Walker universe with a constant expansion
  rate $(w = -1/3).$  Finally, we present a String theory
  background by lifting our six dimensional cosmological solution to
  ten dimensions.
\end{abstract}

\maketitle

\section{General Idea}

The mechanism involved in generating a very small cosmological
constant that satisfies 't~Hooft naturalness is one of the most
pressing questions in contemporary physics. Recent observations of
distant Type Ia supernovae~\cite{Riess:1998cb} strongly indicate that
the universe is expanding in an accelerating phase, with an effective
de-Sitter (dS) constant $H$ that nearly saturates the upper bound
given by the present-day value of the Hubble constant, i.e., $H \alt
H_0 \sim 10^{-33}$~eV. According to the Einstein field equations, $H$
provides a measure of the scalar curvature of the space and is related
to the vacuum energy density $\rho_{\rm vac}$ through Friedmann's
equation, $3\, M_{\rm Pl}^2 H^2 \sim \rho_{\rm vac},$
where $M_{\rm Pl} \simeq 2.4 \times 10^{18}~{\rm GeV}$ is the reduced
Planck mass.  However, the ``natural'' value of $\rho_{\rm vac}$ coming
from the zero-point energies of known elementary particles is found to
be at least $\rho_{\rm vac} \sim {\rm TeV}^4.$
Substitution of this value of $\rho_{\rm vac}$ into
Friedmann's equation yields $H \agt 10^{-3}$~eV, grossly inconsistent
with the set of supernova (SN) observations. The absence of a mechanism in
agreement with 't~Hooft naturalness criteria then centers on the
following question: why is the vacuum energy needed by the
Einstein field equations 120 orders of magnitude smaller than any
``natural'' cut-off scale in effective field theory of particle
interactions, but not zero?

Nowadays, the most popular framework which can address aspects of
this question is the anthropic approach, in which the fundamental
constants are not determined through fundamental reasons, but rather
because such values are necessary for life (and hence intelligent
observers to measure the constants)~\cite{Weinberg:dv}. Of course, in
order to implement this idea in a concrete physical theory, it is
necessary to postulate a multiverse in which fundamental physical
parameters can take different values. Recent investigations in String
theory have applied a statistical approach to the enormous
``landscape'' of metastable vacua present in the
theory~\cite{Bousso:2000xa}.  A vast ensemble of metastable vacua
with a small positive effective cosmological constant that can
accommodate the low energy effective field theory of the Standard Model (SM)
have been found. Therefore, the idea of a string landscape has been
used to proposed a concrete implementation of the anthropic principle.

Nevertheless, the compactification of a String/M-theory background to
a four dimensional solution undergoing accelerating expansion has
proved to be exceedingly difficult. The obstruction to finding dS
solutions in the low energy equations of String/M theory is well
known and summarized in the no-go theorem of~\cite{Maldacena:2000mw}.
This theorem states that in a $D$-dimensional theory of gravity, in
which $(a)$ the action is linear in the Ricci scalar curvature $(b)$
the potential for the matter fields is non-positive and $(c)$ the
massless fields have positive defined kinetic terms, there are no
(dynamical) compactifications of the form: $ds^2_D = \Omega^2(y)
(dx_d^2 + \hat g_{mn} dy^n dy^m)$, if the $d$ dimensional space has
Minkowski $SO(1,d-1)$ or dS $SO(1,d)$ isometries and its $d$
dimensional gravitational constant is finite (i.e., the internal space
has finite volume).  The conclusions of the theorem can be
circumvented if some of its hypotheses are not satisfied. Examples
where the hypotheses can be relaxed exist: $(i)$ one can find
solutions in which not all of the internal dimensions are
compact~\cite{Gibbons:2001wy}; $(ii)$ one may try to find a solution
breaking Minkowski or de Sitter invariance~\cite{Townsend:2003fx};
$(iii)$ one may try to add negative tension matter (e.g., in the form
of orientifold planes)~\cite{Giddings:2001yu}; $(iv)$ one can even
appeal to some intrincate String dynamics~\cite{Kachru:2003aw}.

Salam-Sezgin six dimensional supergravity model~\cite{Salam:1984cj}
provides a specific example where the no-go theorem is not at work,
because when their model is lifted to M theory the internal space is
found to be non-compact~\cite{Cvetic:2003xr}. The lower dimensional
perspective of this, is that in six dimensions the potential can be
positive.  This model has perhaps attracted the most attention because
of the wide range of its phenomenological
applications~\cite{Halliwell:1986bs}. In this article we examine the
cosmological implications of such a supergravity model during the
epochs subsequent to primordial nucleosynthesis. We derive a solution
of Einstein field equations which is in qualitative agreement with
luminosity distance measurements of Type Ia
supernovae~\cite{Riess:1998cb}, primordial nucleosynthesis
abundances~\cite{Olive:1999ij}, data from the Sloan Digital Sky Survey
(SDSS)~\cite{Tegmark:2003ud}, and the most recent measurements from
the Wilkinson Microwave Anisotropy Probe (WMAP)
satellite~\cite{Spergel:2006hy}. The observed acceleration of the
universe is driven by the ``dark energy'' associated to a scalar field
slowly rolling down its exponential potential (i.e., kinetic energy
density $<$ potential energy density $\equiv$ negative
pressure)~\cite{Halliwell:1986ja}. Very interestingly, the resulting
cosmological model also predicts a cold dark matter (CDM) candidate.
In analogy with the phenomenological proposal
of~\cite{Comelli:2003cv}, such a nonbaryonic matter interacts with the
dark energy field and therefore the mass of the CDM particles evolves
with the exponential dark energy potential.  However, an attempt to
saturate the present CDM component in this manner leads to gross
deviations from present cosmological data. We will show that this type
of CDM can account for up to about 7\% of the total CDM budget.
Generalizations of our scenario (using supergravities with more
fields) might account for the rest.

\section{Salam-Sezgin Cosmology}

We begin with the action of Salam-Sezgin six dimensional
supergravity~\cite{Salam:1984cj}, setting to zero the fermionic terms
in the background (of course fermionic excitations will arise from
fluctuations),
\beq S=\frac{1}{4 \kappa^2}\int d^6x \sqrt{g_6}\Big[ R
- \kappa^2 (\partial_M\sigma)^2 -\kappa^2 e^{\kappa\sigma}F_{MN}^2
-\frac{2g^2}{\kappa^2}e^{-\kappa\sigma}
-\frac{\kappa^2}{3}e^{2\kappa\sigma} G_{MNP}^2 \Big] \,\, .
\label{ss}
\eeq
Here, $g_6=\det g_{MN},$ $R$ is the Ricci scalar of $g_{MN},$
$F_{MN}=\partial_{[M} A_{N]},$ $G_{MNP}=\partial_{[M}B_{NP]}
+\k A_{[M} F_{NP]},$ and capital Latin indices run from 0 to 5.
A re-scaling of the constants:
$G_6\equiv2 \kappa^2,$ $\phi \equiv -\k\sigma$ and $\xi\equiv 4\,g^2$
leads to
\beq
S=\frac{1}{2 G_6} \int d^6x\sqrt{g_6}\Big[R -  (\partial_M\phi)^2 -
\frac{\xi}{G_6} e^\phi - \frac{G_6}{2} e^{-\phi} F_{MN}^2 -
\frac{G_6}{6}
e^{-2\phi} G_{MNP}^2\Big] \,\,.
\label{action}
\eeq
The length dimensions of the
fields are: $[G_6]=L^4,$ $[\xi]=L^2,$ $[\phi]=[g_{MN}^2]=1,$
$[A_M^2]=L^{-4},$ and $[F_{MN}^2]=[G_{MNP}^2]= L^{-6}.$

Now, we consider a
spontaneous compactification from six dimension to four dimension.
To this end, we take the six dimensional manifold $M$ to be
a direct product of 4 Minkowski directions (hereafter denoted by
$N_1$) and a compact orientable two dimensional manifold $N_2$ with
constant curvature.  Without loss of generality, we can set
$N_2$ to be a sphere $S^2$, or a $\Sigma_2$ hyperbolic manifold
with arbitrary genus. The metric on $M$ locally takes the form
\bea ds_6^2= ds_4(t,{\vec x})^2 + e^{2f(t,{\vec x})}d\sigma^2, &&
d\sigma^2=\left\{\begin{array}{ll}
   r_c^2\, (d\vartheta^2 +\sin^2\vartheta d\varphi^2) & \,\, {\rm for} \ S^2 \\
   r_c^2\, (d\vartheta^2 +\sinh^2\vartheta d\varphi^2) & \,\, {\rm for}\
\Sigma_2 \,\,,
  \end{array} \right.
\label{metric}
\eea
where $(t,\ \vec x)$ denotes a local coordinate system in $N_1,$ $r_c$ is
the compactification radius of $N_2$.
We assume that the scalar field $\phi$ is only
dependent on the point of $N_1$, i.e., $\phi=\phi(t,\ \vec x)$. We further
assume that the gauge field $A_M$ is excited on $N_2$ and is of the form
\bea
A_\varphi=\left\{\begin{array}{ll}
    b\cos \vartheta &(S^2) \\
    b\cosh \vartheta& (\Sigma_2) \, .
  \end{array} \right. \label{FC}
\eea
This is the monopole configuration detailed by
Salam-Sezgin~\cite{Salam:1984cj}. Since
we set the Kalb-Ramond field $B_{NP} = 0$ and the term
$A_{[M} F_{NP]}$ vanishes on $N_2$, $G_{MNP} = 0$.
The field strength becomes
\begin{equation}
F_{MN}^2=
    2b^2 e^{-4f}/r_c^4  \, .
\label{FS}
\end{equation}
Taking the variation of the gauge field $A_M$ in Eq.~(\ref{action})
we obtain the Maxwell equation
\be
\partial_M \Big[ \sqrt{g_4}\sqrt{g_\sigma} e^{2f-\phi} F^{MN}  \Big]=0.
\label{Maxwell}
\ee
It is easily seen that the field strengths in Eq.~(\ref{FS}) satisfy
Eq.~(\ref{Maxwell}).

With this in mind, the Ricci scalar reduces to~\cite{Wald:1984rg} \be
R[M]=R[N_1]+e^{-2f}R[N_2]-4\Box f-6(\partial_\m f)^2 \,\,, \ee where
$R[M],$ $R[N_1],$ and $ R[N_2]$ denote the Ricci scalars of the
manifolds $M,$ $N_1,$ and $N_2$; respectively. (Greek indices run from
0 to 3). The Ricci scalar of $N_2$ reads \bea R[N_2]=
\left\{\begin{array}{ll}
    +2/r_c^2 & (S^2)\\
    -2/r_c^2 & (\Sigma_2).
\end{array} \right.
\label{007}
\eea
To simplify the notation, from now on, $R_1$ and $R_2$ indicate $R[N_1]$
and $R[N_2]$, respectively. The determinant of the metric can be written as
$\sqrt{g_6}=e^{2f}\sqrt{g_4}\sqrt{g_\sigma},$
where $g_4=\det g_{\m\n}$ and $g_\sigma$ is the determinant of the metric
of $N_2$ excluding the factor $e^{2f}$.
We define the gravitational constant in the four dimension as
\be
\frac{1}{G_4}\equiv \frac{M_{\rm Pl}^2}{2} = \frac{1}{2\, G_6}
\int d^2\sigma \sqrt{g_{\sigma}} =
\frac{2 \pi r_c^2}{G_6} \,\, .
\ee
Hence, by using the field configuration given in Eq.~(\ref{FC}) we
can re-write the action in Eq.~(\ref{action}) as follows
\be
S= \frac{1}{G_4} \int d^4 x \sqrt{g_4} \Big\{ e^{2f}\big[ R_1 +
e^{-2f} R_2 + 2 (\partial_\mu f)^2 - (\partial_\mu \phi)^2 \big]  -
\frac{\xi}{G_6} e^{2f+\phi} - \frac{G_6 b^2}{r_c^4}\, e^{ -2f-\phi}
\Big\} \,.
\label{action4string}
\ee
Let us consider now a rescaling of the metric of $N_1$:
$\hat{g}_{\mu\nu}\equiv e^{2f} g_{\mu\nu}$ and
$\sqrt{\hat{g}_4}=e^{4f} \sqrt{g_4}.$ Such a transformation brings
the theory into the Einstein conformal frame where the action given in
Eq.~(\ref{action4string}) takes the form
\be S= \frac{1}{G_4} \int d^4
x \sqrt{\hat{g}_4} \Big[ R[\hat{g}_4] -4 (\partial_\mu f)^2 -
(\partial_\mu \phi)^2 - \frac{\xi}{G_6} e^{-2f+\phi}-\frac{G_6 b^2}{r_c^4}\,
e^{-6f-\phi} + e^{-4f}R_2 \Big].
\label{action4}
\ee
The four dimensional Lagrangian is then
\beq
L= \frac{\sqrt{g}}{G_4}\,\Big[ R- 4 (\partial_\mu f)^2 - (\partial_\mu
\phi)^2 - V(f,\phi)  \Big],
\label{finallagran}
\eeq
with
\beq
V(f,\phi)\equiv \frac{\xi}{G_6}e^{-2f+\phi} + \frac{G_6 b^2}{r_c^4}\,
e^{-6f-\phi} - e^{-4f}R_2 \,\,,
\eeq
where to simplify the notation we have defined: $g\equiv\hat{g}_4$
and $R\equiv R[\hat{g}_4]$.

Let us now define a new orthogonal basis, $X \equiv (\phi+2f)/\sqrt G_4$
and $Y \equiv (\phi-2f)/\sqrt G_4$, so that the kinetic energy terms in
the Lagrangian are both canonical, i.e.,
\beq
L= \sqrt{g}\left[\frac{R}{G_4} -\frac{1}{2}(\partial X)^2
-\frac{1}{2}(\partial Y)^2 - \tilde V(X,Y)  \right],
\label{ll}
\eeq
where the potential $\tilde V(X,Y) \equiv V( f, \phi)/G_4$ can be re-written (after some elementary algebra) as~\cite{Vinet:2005dg}
\beq
\tilde V(X,Y)= \frac{e^{\sqrt{G_4} Y}}{G_4} \left[ \frac{G_6
b^2}{r_c^4}e^{-2\sqrt{G_4} X} - R_2 e^{-\sqrt{G_4} X} +\frac{\xi}{G_6}
\right] \,\, .
\label{potential}
\eeq

The field equations are
\bea
 R_{\mu\nu} - \frac{1}{2}g_{\mu\nu}R & = &
    \frac{G_4}{2} \left[\left(\partial_\mu X \partial_\nu X -\frac{g_{\mu\nu}}{2}\, \partial_\eta X \, \partial^\eta X \right) \right. \nonumber \\
 & + &
\left. \left(
\partial_\mu Y \partial_\nu Y -\frac{g_{\mu\nu}}{2} \,\partial_\eta Y \, \partial^\eta Y \right) -g_{\mu\nu} \tilde V(X,Y)\right] \, ,
\label{einsteinb}
\eea
$\Box  X = \partial_X \tilde V,$ and $\Box  Y = \partial_Y \tilde V.$
In order to allow for a dS era we assume that the metric
takes the form
\beq
ds^2= -dt^2 + e^{2h(t)}d\vec{x}^{\,2}  ,
\label{refc}
\eeq
and that $X$ and $Y$ depend only on the time coordinate, i.e., $X=X(t)$
and $Y=Y(t)$.
Then the equations of motion for $X$ and $Y$ can be written as
\bea
\ddot X + 3 \dot h\dot X=  - \partial_X \tilde V
\eea
and
\begin{equation}
 \ddot Y + 3 \dot h\dot Y= - \partial_Y \tilde V,
\label{eqy}
\end{equation}
whereas the  only two independent components
of Eq.~(\ref{einsteinb}) are
\begin{equation}
\dot h^2 = \frac{G_4}{6}\left[\frac{1}{2}(\dot X^2 + \dot Y^2) + \tilde V(X,Y)\right]
\label{hubble}
\end{equation}
and
\begin{equation}
 2 \ddot h+3\dot h^2 =  \frac{G_4}{2}\left[-\frac{1}{2}( \dot X^2 +  \dot Y^2) + \tilde V(X,Y)\right]\, .
\end{equation}

The terms in the square brackets in Eq.~(\ref{potential}) take the
form of a quadratic function of $e^{-\sqrt{G_4}\,X}.$ This function
has a global minimum at $ e^{-\sqrt{G_4}\, X_0} = R_2\, r_c^4/(2\,
G_6\,b^2).$ Indeed, the necessary and sufficient condition for a
minimum is that $R_2 >0$, so hereafter we only consider the spherical
compactification, where $e^{-\sqrt{G_4} \, X_0} = M_{\rm Pl}^2 /(4 \pi
b^2).$ The condition for the potential to show a dS rather than an AdS
or Minkowski phase is $\xi b^2 > 1$. Now, we expand
Eq.~(\ref{potential}) around the minimum,
\begin{equation}
\tilde V(X,Y) =  \frac{e^{\sqrt{G_4}\, Y}}{G_4} \,  \left[
    {\cal K} + \frac{\overline{M_X}^2}{2}  (X-X_0)^2 +
    {\cal O} \Big((X- X_0)^3 \Big)\right] \,,
\label{minBUF}
\end{equation}
where
\begin{equation}
\overline{M_X} \equiv \frac{1}{\sqrt{\pi}\,\,\, b r_c}
\label{mxbar}
\end{equation}
and
\begin{equation}
{\cal K} \equiv \frac{M_{\rm Pl}^2}{4 \pi r_c^2 b^2} (b^2 \xi - 1) \,\, .
\label{calk}
\end{equation}
As shown by Salam-Sezgin~\cite{Salam:1984cj} the requirements for
preserving a fraction of supersymmetry (SUSY) in spherical
compactifications to four dimension imply $b^2 \xi = 1$, corresponding
to winding number $n= \pm 1$ for the monopole configuration.
Consequently, a ($Y$-dependent) dS background can be obtained only
through SUSY breaking. For now we will leave open the symmetry
breaking mechanism and come back to this point after our
phenomenological discussion. The $Y$-dependent physical mass of the
$X$-particles at any time is
\begin{equation}
M_X (Y)= \frac{e^{\sqrt{G_4}\, Y/2}}{\sqrt{G_4}}\ \overline{M_X}\,,
\label{mphys}
\end{equation}
which makes this a varying mass particle (VAMP)
model~\cite{Comelli:2003cv}, although, in this case, the dependence on
the quintessence field is fixed by the theory.
The dS (vacuum) potential energy density is
\begin{equation}
V_Y = \frac{e^{\sqrt{G_4}\, Y}}{G_4}\ {\cal K} \, .
\label{rowvac}
\end{equation}
In general,
classical oscillations for the $X$ particle will occur for
\begin{equation}
M_X > H = \sqrt{\frac{G_4\rho_{\rm tot}}{3}} \,,
\label{osc}
\end{equation}
where $\rho_{\rm tot}$ is the total energy density. (This condition is well
known from axion cosmology~\cite{Preskill:1982cy}).  A necessary
condition for this to hold  can be obtained by saturating
$\rho$ with $V_{Y}$ from
Eq.~(\ref{rowvac}) and making use of Eqs.~(\ref{mxbar}) to
(\ref{osc}), which leads to $\xi b^2 <7$. Of course, as we stray from
the present into an era where the dS energy is not dominant, we
must check at every step whether the inequality (\ref{osc}) holds. If
the inequality is violated, the $X$-particle ceases to behave like
CDM.

In what follows, some combination of the parameters of the model will
be determined by fitting present cosmological data.    To
this end we assume that SM fields are confined to $N_1$ and we
denote with $\rho_{\rm rad}$ the radiation energy, with
$\rho_X$ the matter energy associated with the $X$-particles, and with
$\rho_{\rm mat}$ the remaining matter density. With this in mind,
Eq.~(\ref{eqy}) can be re-written as \be \ddot Y + 3 \,H \,\dot Y = -
\frac{\partial V_{\rm eff}}{\partial Y} \,, \ee where $V_{\rm eff}
\equiv V_Y +\rho_X$ and $H$ is defined by the Friedmann equation \be
H^2 \equiv \dot{h}^2 = \frac{1}{3 M_{\rm Pl}^2} \, \left[
  \frac{1}{2}\, \dot Y^2 + V_{\rm eff} + \rho_{\rm rad} + \rho_{\rm
    mat} \right] \,\,.  \ee (Note that the matter energy associated to
the $X$ particles is contained in $V_{\rm eff}$.)

It is more convenient to consider the evolution in $u \equiv - \ln (1 +z),$
where $z$ is the redshift parameter.
As long as the oscillation condition is fulfilled, the VAMP CDM energy
density is given in terms of the $X$-particle number density
$n_X$~\cite{Hoffman:2003ru}
\begin{equation}
\rho_X(Y,u) = M_X(Y)\ n_X(u)\\[.1in]
= C \ e^{\sqrt{G_4} Y/2}\ e^{-3u} \,\,,
\label{rowx}
\end{equation}
where $C$ is a constant to be determined by fitting to data.
Along with Eq.~(\ref{rowvac}), these define for us the effective
($u$-dependent) VAMP potential
\be V_{\rm eff}(Y,u)\equiv V_Y +\rho_X=  A\ e^{\sqrt{G_4}Y} +
C \
e^{\sqrt{G_4}Y/2}\ e^{-3u}\, ,
\label{veff}
\ee
where a $A$ is just a constant given in terms of model parameters through
Eqs.~(\ref{minBUF}) and (\ref{calk}).

Hereafter we adopt natural units, $M_{\rm Pl} = 1.$ Denoting by a
prime derivatives with respect to $u,$ the equation of motion for $Y$
becomes \be \frac{Y^{\prime\prime}}{1 - \,
  Y^{\prime^2}\!/6} \,\, + 3 \,Y^\prime +
\frac{\partial_u\rho\,\, \,Y^\prime/2 \,\,+ \,3 \,\,\partial_Y V_{\rm
    eff}}{\rho} = 0\,,
\label{motion}
\ee
where $\rho = V_{\rm eff} + \rho_{\rm rad} + \rho_{\rm mat}.$
Quantities of importance are the dark energy density
\be \rho_{Y} =
\frac{1}{2}\, H^2 \,Y^{\prime 2} + V_Y \,,
\ee
generally expressed in
units of the critical density ($\Omega\equiv\rho/\rho_{\rm c}$)
\be
\Omega_Y = \frac{\rho_Y}{3 H^2}\,,
\ee
and the Hubble parameter
\be
H^2 = \frac{\rho}{3 - Y^{\prime 2}/2}\,\,.
\ee
The equation of state
is
\be w_Y = \left[\frac{H^2 \,\,Y^{\prime 2}}{2} - V_Y\right]\,
\left[\frac{H^2 \,\,Y^{\prime 2}}{2}+ V_Y\right]^{-1}\,.
\ee
We pause to note that the exponential potential $V_Y
\sim e^{\lambda Y/M_{\rm Pl}},$ with $\lambda =\sqrt{2}.$
Asymptotically,
this represents the crossover situation with $w_Y = -1/3$~\cite{copeland},
implying expansion at constant velocity. Nevertheless, we will find that there
is a brief period encompassing the recent past $(z\alt 6)$ where there has
been significant acceleration.

Returning now to the quantitative analysis, we take
$\rho_{\rm mat} = B e^{-3u}$ and $\rho_{\rm rad} = 10^{-4} \
\rho_{\rm mat} \ e^{-u}\ f(u)$~\cite{note} where $B$ is a constant and
$f(u)$ parameterizes the $u$-dependent number of radiation degrees of
freedom.  In order to interpolate the various thresholds appearing
prior to recombination (among others, QCD and electroweak), we adopt a
convenient phenomenological form
$f(u)=\exp(-u/15)$~\cite{Anchordoqui:2003ij}.  We note at this point
that solutions of Eq.~(\ref{motion}) are independent by an overall
normalization for the energy density. This is also true for the
dimensionless quantities of interest $\Omega_Y$ and $w_Y.$

\begin{figure}[!thb]
\postscript{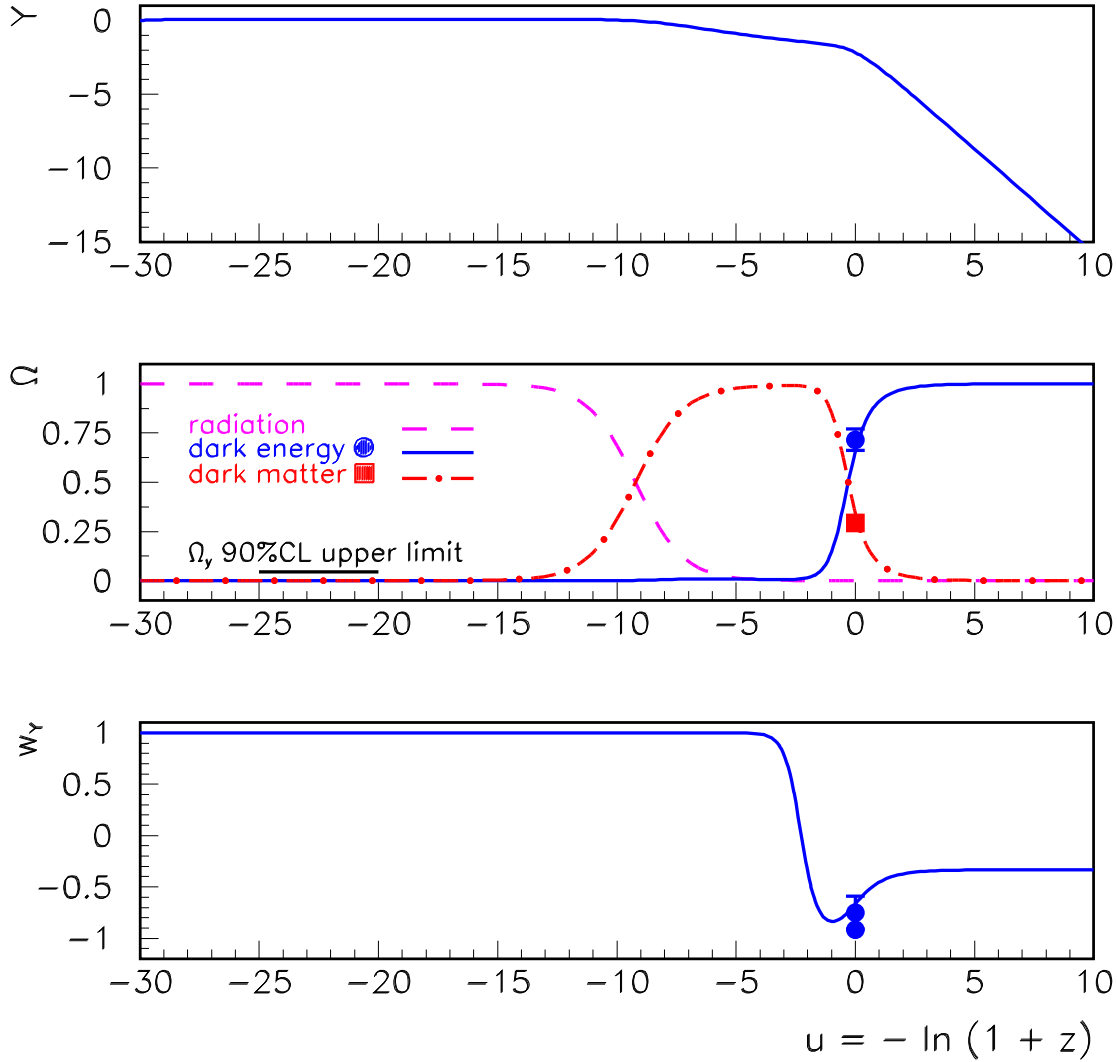}{0.9}
\caption{The upper panel shows the evolution of $Y$ as a function of
  $u$. Today corresponds to $z=0$ and for primordial nucleosynthesis
  $z \approx 10^{10}.$ We set the initial conditions $Y (-30) = 0$ and
  $Y'(-30) = 0.08;$ we take $A : B : C = 11 : 0.3 : 0.1.$
  The second panel shows the evolution of
  $\Omega_Y$ (solid line), $\Omega_{\rm mat}$ (dot-dashed line), and
  $\Omega_{\rm rad}$ (dashed line) superposed over experimental best
  fits from SDSS and WMAP
  observations~\cite{Tegmark:2003ud,Spergel:2006hy}. The curves are
  not actual fits to the experimental data but are based on the
  particular choice of the $Y$ evolution shown in the upper panel,
  which provides eyeball agreement with existing astrophysical
  observations.  The lower panel shows the evolution of the equation
  of state $w_Y$ superposed over the best fits to WMAP + SDSS data
  sets and WMAP + SNGold~\cite{Spergel:2006hy} . The solution of the
  field equations is consistent with the requirement from primordial
  nucleosynthesis, $\Omega_Y < 0.045$ (90\%CL)~\cite{Olive:1999ij}, it
  also shows the established radiation and matter dominated epochs,
  and at the end shows an accelerated dS era.}
\label{sancle}
\end{figure}

With these forms for the energy densities, Eq.~(\ref{motion}) can be
integrated for various choices of $A,$ $B,$ and $C$, and initial
conditions at $u=-30.$ We take as initial condition $Y(-30) = 0$.
Because of the slow variation of $Y$ over the range of $u,$ changes in
$Y(-30)$ are equivalent to altering the quantities $A$ and
$C$~\cite{LopesFranca:2002ek}. In accordance to equipartition
arguments~\cite{LopesFranca:2002ek,Steinhardt:1999nw} we take $Y'(-30)
= 0.08.$ Because the $Y$ evolution equation depends only on energy
density ratios, and hence only on the ratios $A:B:C$ of the previously
introduced constants, we may, for the purposes of integration and
without loss of generality, arbitrarily fix $B$ and then scan the $A$
and $C$ parameter space for applicable solutions. In Fig.~\ref{sancle}
we show a sample qualitative fit to the data. It has the property of
allowing the maximum value of $X$-CDM (about 7\% of the total dark
matter component) before the fits deviate unacceptably from data.

It is worth pausing at this juncture to examine the consequences of
this model for variation in the fine structure constant and long range
forces. Specifically, excitations of the electromagnetic field on
$N_1$ will, through the presence of the dilaton factor in
Eq.~(\ref{action}), seemingly induce variation in the electromagnetic
fine structure constant $\alpha_{\rm em} = e^2/4\pi$, as well as a
violation of the equivalence principle through a long range coupling
of the dilaton to the electromagnetic component of the stress tensor.
We now show that these effects are extremely negligible in the present model.
First, it is easily seen using Eqs.~(\ref{action}) and (\ref{metric})
together with Eqs.~(\ref{007})-(\ref{potential}), that the
electromagnetic piece of the lagrangian as viewed from $N_1$ is 
\begin{equation}
 {\cal L}_{\rm em} = -\frac{2\pi}{4} e^{- {\sqrt{G_4}X}} 
\widetilde{f}_{\mu\nu}^2 \,\,,
\end{equation}
where $\widetilde{f}_{\mu \nu}$ denotes a quantum fluctuation of the
electromagnetic $U(1)$ field. (Fluctuations of the $U(1)$ background
field are studied in the Appendix). At the equilibrium value
$X=X_0$, the exponential factor is
\begin{equation}
 e^{- {\sqrt{G_4}X_0}} =
 \frac{M_{\rm Pl}^2}{4\pi b^2} \,,
\end{equation}
so that we can identify the electromagnetic coupling $(1/e^2) \simeq
M_{\rm Pl}^2/ b^2$. This shows that $b\sim M_{\rm Pl}$.  We can then
expand about the equilibrium point, and obtain an additional factor of
$(X-X_0)/M_{\rm Pl}$. This will do two things~\cite{Carroll:1998zi}:
$(a)$ At the classical level, it will induce a variation of the
electromagnetic coupling as $X$ varies, with $\Delta\alpha_{\rm
  em}/\alpha_{\rm em} \simeq (X-X_0)/M_{\rm Pl}$; $(b)$~at the quantum
level, exchange of $X$ quanta will induce a new force through coupling
to the electromagnetic component of matter.

Item $(b)$ is dangerous if the mass of the exchanged quanta are small,
so that the force is long range.  This is not the case in the present
model: from Eq.~(\ref{minBUF}) the $X$ quanta have mass of ${\cal
  O}({\overline M}_X M_{\rm Pl}) \sim M_{\rm Pl}/(r_c b)$, so that if
$r_c$ is much less than ${\cal O}({\rm cm}),$ the forces will play
no role in the laboratory or cosmologically.

As far as the variation of $\alpha_{\rm em}$ is concerned, we find that
$\rho_X/\rho_{\rm mat} = (C/B) e^{Y/\sqrt{2}},$ so that
\begin{eqnarray}
\rho_X &\simeq& 3\times 10^{-120} e^{-3u}M_{\rm Pl}^4 e^{Y/\sqrt{2}}
\nonumber \\
&=& \frac{1}{4} \overline M _X^2 (X-X_0)^2 e^{Y \sqrt{2}}M_{\rm Pl}^2 \,\,.
\end{eqnarray}
This then gives,
\begin{equation}
 \sqrt{\langle(X-X_0)^2 \rangle }\equiv \Delta X_{\rm rms}  \approx 10^{-60}
e^{-3u/2} M_{\rm Pl} e^{Y/(2\sqrt{2})}/\overline M _X \,\, .
\end{equation}
During the radiation era, $Y\simeq \ {\rm const} \ \simeq 0$ (see
Fig.~\ref{sancle}), so that during nucleosynthesis ($u\simeq -23)$
$\Delta X_{\rm rms}/M_{\rm Pl} \simeq 10^{-45}/\overline M _X,$
certainly no threat. It is interesting that such a small value can be
understood as a result of inflation: from the equation of motion for
the $X$ field, it is simple to see that during a dS era with
Hubble constant $H$, the amplitude $\Delta X_{\rm rms}$ is damped as
$e^{-3Ht/2}$. For 50~$e$-foldings, this represents a damping of
$10^{32}.$ In order to make the numbers match (assuming a
pre-inflation value $\Delta X_{\rm rms}/M_{\rm Pl}\sim 1$) an
additional damping of $\sim 10^{13}$ is required from reheat
temperature to primordial
nucleosynthesis. With the $e^{-3u/2}$ behavior, this implies a low
reheat temperature, about $10^6$ GeV.
Otherwise, one may just assume an additional fine-tuning of the
initial condition on $X$.

As mentioned previously, the solutions of Eq.~(\ref{motion}), as well
as the quantities we are fitting to ($\Omega_Y$ and $w_Y$), depend
only on the ratios of the energy densities. From the eyeball fit in
Fig.~\ref{sancle} we have, up to a common constant, $\rho_{\rm
  ordinary\ matter} \equiv \rho_{\rm mat} \propto 0.3\ e^{-3u}$ and $
V_Y \propto 11\ e^{\sqrt{2}Y}.$ We can deduce from these relations
that \be \frac{V_Y({\rm now})}{\rho_{\rm mat}({\rm now})} =
\frac{11}{0.3}\ e^{\sqrt{2}Y({\rm now})} \simeq 36\ e^{\sqrt{2}Y({\rm
    now})} \,\,.
\label{ratio}
\ee
Besides, we know that $\rho_{\rm mat}({\rm now}) \simeq 0.3
\rho_c({\rm now})\simeq 10^{-120}\
M_{\rm Pl}^4.$ Now, Eqs.~(\ref{minBUF}) and (\ref{calk}) lead to
\be
 V_Y({\rm now}) = e^{\sqrt{2}Y({\rm now})}\
\frac{M_{\rm Pl}^4}{8\pi\ r_c^2\ b^2}\ (b^2\xi -1)
\label{theory}
\ee
so that from Eqs.~(\ref{ratio}) and (\ref{theory}) we obtain
\be
\frac{1}{8\pi\ r_c^2\ b^2}\ (b^2\xi -1) \simeq 10^{-119} \,\,.
\label{finetuning}
\ee It is apparent that this condition cannot be naturally
accomplished by choosing large values of $r_c$ and/or $b.$ There
remains the possibility that SUSY breaking~\cite{Aghababaie:2003wz} or
non-perturbative effects lead to an exponentially small deviation of
$b^2 \xi$ from unity, such that $b^2 \xi = 1 + {\cal O}
(10^{-119})$~\cite{cstring}. Since a deviation of $b^2\xi$ from
unity involves a breaking of supersymmetry, a small value for this
dimensionless parameter, perhaps $(1\ {\rm TeV}/M_{\rm Pl})^2\sim
10^{-31}$, can be expected on the basis of 't Hooft naturalness.  It
is the extent of the smallness, of course, which remains to be
explained.

\section{The String Connection}

We now briefly comment on how the six dimensional solution derived
above reads in String theory. To this end, we use the uplifting
formulae developed by Cvetic, Gibbons and Pope~\cite{Cvetic:2003xr};
we will denote with the subscript ``cgp'' the quantities of that paper
and with ``us'' quantities in our paper.  Let us more specifically
look at Eq.~(34) in Ref.~\cite{Cvetic:2003xr}, where the authors
described the six dimensional Lagrangian they uplifted to Type I
String theory. By simple inspection, we can see that the relation
between their variables and fields with the ones we used in
Eq.~(\ref{action}) is $ \phi|_{\rm cgp}= -2\phi|_{\rm us},$ $F_2|_{\rm
  cgp} = \sqrt{G_6} F_2|_{\rm us},$ $H_3|_{\rm cgp} =\sqrt{G_6/3}\,
G_{3}|_{\rm us},$ and $\bar{g}^2|_{\rm cgp} = \xi/(8 G_6)|_{\rm us}.$
Our six dimensional background is determined by the (string frame)
metric $ ds_6^2 = e^{2f}\, \Big[- dt^2 + e^{2h}dx_3^2 + r_c^2 \,
d\sigma_2{}^2 \Big],$ the gauge field $F_{\vartheta \varphi}=-b \sin
\vartheta,$ and the $t$-dependent functions $h(t),$ $f(t)=
\sqrt{G_4}\, (X-Y)/4,$ and $\phi(t)=\sqrt{G_4}\, (X+Y)/2.$ Identifying
these expressions with those in Eqs.~(47), (48) and (49) of
Ref.~\cite{Cvetic:2003xr} one obtains a full Type I or Type IIB
configuration, consisting of a 3-form (denoted by $F_3$),
\begin{eqnarray}
F_3 & = & \frac{8 G_6 \sinh\hat \rho \cosh\hat \rho}{\xi \cosh^2 2\hat \rho}d\hat \rho \wedge
\Big(d\alpha -\sqrt{\frac{\xi}{8 G_6}} b
\cos\vartheta d\varphi \Big) \wedge \Big(d\beta +\sqrt{\frac{\xi}{8 G_6}} b
\cos\vartheta d\varphi \Big) \nonumber \\
& - & \frac{\sqrt{2} G_6 b}{\sqrt{\xi} \cosh 2\hat \rho} \sin\vartheta d\theta
\wedge
d\varphi \wedge \left[ \cosh^2\hat\rho \left(d\alpha -\sqrt{\frac{\xi}{8 G_6}} b
\cos\vartheta d\varphi \right) \right. \nonumber \\
& - & \left. \sinh^2\hat \rho \left(d\beta
+\sqrt{\frac{\xi}{8 G_6}} b
\cos\vartheta d\varphi \right)  \right] \,,
\end{eqnarray}
a  dilaton (denoted by $\hat{\phi}$)
\begin{equation}
e^{2\hat{\phi}}=\frac{e^{2\phi}}{\cosh(2\hat \rho)} \,,
\end{equation}
and a ten dimensional metric that in the
string frame
reads
\begin{eqnarray}
ds^2_{\rm str} &=& e^{\phi}\, ds_6^2 + dz^2 +\frac{4G_6}{\xi} \left[ d\hat \rho^2
+\frac{\cosh^2\hat \rho}{\cosh 2\hat \rho} \left( d\alpha -\sqrt{\frac{\xi}{8 G_6}} b
\cos\vartheta d\varphi \right)^2 \right. \nonumber \\
 & + & \left. \frac{\sinh^2\hat \rho}{\cosh 2\hat \rho}
\left(d\beta +\sqrt{\frac{\xi}{8 G_6}} b
\cos\vartheta d\varphi \right)^2\right] \,,
\label{config10d}
\end{eqnarray}
where $\hat \rho,\, z,\, \alpha,$ and $\beta$ denote the four extra
coordinates.  It is important to stress that though the uplifted
procedure decribed above implies a non-compact internal manifold, the
metric in Eq.~(\ref{config10d}) can be interpreted within the context
of~\cite{Giddings:2001yu} (i.e., $0 \leq \hat \rho \leq L,$ with
$L \gg 1$ an infrared cutoff where the spacetime smoothly closes up)
to obtain a finite volume for the internal space and consequently a
non-zero but tiny value for $G_6.$

\section{Conclusions}

We studied the six dimensional Salam-Sezgin model~\cite{Salam:1984cj},
where a solution of the form Minkowski$_{4} \times S^2$ is known to
exist, with a $U(1)$ monopole serving as background in the two-sphere.
This model circumvents the hypotheses of the no-go
theorem~\cite{Maldacena:2000mw} and then when lifted to String theory
can show a dS phase. In this work we have allowed for time dependence
of the six-dimensional moduli fields and metric (with a
Robertson-Walker form).  Time dependence in these fields vitiates
invariance under the supersymmetry transformations.  With these
constructs, we have obtained the following results:

\begin{flushleft}

  (1) In terms of linear combinations of the $S^2$ moduli field and
  the six dimensional dilaton, the effective potential consists of
  $(a)$ a pure exponential function of a quintessence field (this
  piece vanishes in the supersymmetric limit of the static theory) and
  $(b)$ a part which is a source of cold dark matter, with a mass
  proportional to an exponential function of the quintessence field.
  This presence of a VAMP CDM candidate is inherent in the model.
  \smallskip

  (2) If the monopole strength is precisely at the value prescribed by
  supersymmetry, the model is in gross disagreement with present
  cosmological data -- there is no accelerative phase, and the
  contribution of energy from the quintessence field is purely
  kinetic.  However, a miniscule deviation of ${\cal O}(10^{-120}$)
  from this value permits a qualitative match with data. Contribution
  from the VAMP component to the matter energy density can be as large
  as about 7\% without having negative impact on the fit.  The
  emergence of a VAMP CDM candidate as a necessary companion of dark
  energy has been a surprising aspect of the present findings, and
  perhaps encouraging for future exploration of candidates which can
  assume a more prominent role in the CDM sector.  \smallskip

  (3) In our model, the exponential potential $V_Y\sim e^{\lambda
    Y/M_{\rm Pl}}$, with $Y$ the quintessence field and $\lambda =
  \sqrt{2}.$ The asymptotic behavior of the scale factor for
  exponential potentials $e^{h(t)} \approx t^{2/\lambda^2},$ so that
  for our case $h \approx \ln t,$ leading to a conformally flat
  Robertson-Walker metric for large times.  The deviation from
  constant velocity expansion into a brief accelerated phase in the
  neighborhood of our era makes the model phenomenologically viable.
  In the case that the supersymmetry condition ($b^2\xi =1$) is
  imposed, and there is neither radiant energy nor dark matter except
  for the $X$ contribution, we find for large times that the scale
  parameter $e^{h(t)} \approx \sqrt{t},$ so that even in this case the
  asymptotic metric is Robertson-Walker rather than Minkowski.
  Moreover, and rather intriguingly, the scale parameter is what one
  would find with radiation alone~\cite{xcdm}.  \smallskip

\end{flushleft}

  In sum, in spite of the shortcomings of the model (not a perfect
  fit, requirement of a tiny deviation from supersymmetric
  prescription for the monopole embedding), it has provided a
  stimulating new, and unifying, look at the dark energy and dark
  matter puzzles.

  \acknowledgements{We would like to thank Costas Bachas and Roberto
    Emparan for valuable discussions.  The research of HG was
    supported in part by the National Science Foundation under Grant
    No. PHY-0244507.}

\section{Appendix}

In this appendix we study the quantum fluctuations of the $U(1)$ field
associted to the background configuration.  We start by considering
fluctuations of the background field $A_M^0$ in the 4 dimensional
space, i.e,
\begin{equation}
A_M \to A_M^0 + \epsilon \ a_M \,,
\label{amu}
\end{equation}
where $A_M^0 = 0$ if $M \neq \varphi$ and $a_M = 0$ if 
$M = \vartheta, \varphi.$ The fluctuations on $A_M^0$ lead to
\begin{equation}
F_{MN} \to F_{MN}^0 + \epsilon \ f_{M N} \,\, .
\label{fmunu}
\end{equation}
Then,
\begin{equation}
F_{MN} F^{MN} =  g^{ML} \ g^{NP} [F_{MN}^0 F^0_{LP} + \epsilon \ F_{MN}^0 \ f_{LP}  + 
\epsilon^2 f_{MN}\ f_{LP}] \  .
\label{Fdos}
\end{equation}
The second term vanishes and the first and third terms are nonzero
because $F_{MN}^0 \neq 0$ in the compact space and $f_{MN} \neq 0$ in
the 4 dimensional space.  If the Kalb-Ramond potential $B_{NM} = 0$,
then the 3-form field strength can be written as
\begin{equation}
G_{MNP} =\kappa A_{[M} \ F_{NP]}=
\frac{\kappa}{3!}\ [A_M \ F_{NP} + A_P \ F_{MN} - A_N \ F_{MP}]\, .
\end{equation}
Now we introduce notation of differential forms, in which
the usual Maxwell field and field strenght read
\beq
A_1= A_M dx^M\;\; {\rm and} \;\; F_2= F_{MN} \, dx^M \wedge dx^N\,\,;
\eeq
respectively. (Note that $dx^M \wedge dx^N$ is antisymmetrized by 
definition.) With this in mind the 3-form reads
\beq
G_3= \kappa A_1 \wedge F_2= \kappa A_M F_{NP} \  dx^M \wedge dx^N
\wedge dx^P \,\, .
\label{nes}
\eeq
Substituting Eqs.~(\ref{amu}) and (\ref{fmunu}) into Eq.~(\ref{nes}) we obtain
\beq
G_3=\kappa \Big[ (A_M^0 + \epsilon a_M)(F_{NP}^0 + \epsilon f_{N P}) \
dx^M \wedge dx^N \wedge dx^P\Big] \,\, .
\label{1}
\eeq
The background fields read
\beq
A_1^0= b\, \cos\vartheta \, d\varphi,\;\;\; F_2^0= -b\, 
\sin\vartheta\, d\vartheta \wedge d\varphi \,\,,
\label{ocho}
\eeq 
and the fluctuations on the probe brane become
\beq
a_1= a_\mu dx^\mu,\;\;\; f_2= f dx^\mu \wedge dx^\nu,\;\;{\rm with}\;\; f= 
\partial_\mu a_\nu -\partial_\mu a_\nu \, .
\eeq
All in all,
\begin{eqnarray}
\frac{G_3}{\kappa} & = & A^{0}_\varphi F^{0}_{\vartheta\varphi} \ 
d\varphi\wedge 
d\vartheta \wedge d\varphi +\epsilon A_\varphi^0 f_{\mu \nu} \ d\varphi 
\wedge dx^\mu 
\wedge dx^\nu + \epsilon F_{\vartheta\varphi}^0 a_\mu \ d\vartheta\wedge 
d\varphi\wedge 
dx^\mu \nonumber \\
 & + & \epsilon^2 a_\mu f_{\zeta \nu} dx^\mu\wedge dx^\zeta \wedge dx^\nu \ .
\label{2} 
\end{eqnarray}
Using Eq.~(\ref{ocho}) and the antisymmetry of the wedge product,
Eq.~(\ref{2}) can be re-written as \beq
\frac{G_3}{\kappa}=\epsilon\Big[b\cos\vartheta f_{\mu \nu}d\varphi
\wedge dx^\mu \wedge dx^\nu - b a_\mu \sin\vartheta d\vartheta\wedge
d\varphi\wedge dx^\mu +\epsilon a_\mu f_{\zeta \nu} dx^\mu \wedge
dx^\zeta \wedge dx^\nu\Big] \, .
\label{3}
\eeq
From the metric
\beq
ds^2= e^{2\alpha}dx_{4}^2 + e^{2\beta} 
(d\vartheta^2 +\sin\vartheta^2 d\varphi^2)
\eeq
we can write the vielbeins 
\bea
& & e^{a}= e^\alpha dx^a,\;\;\; e^\vartheta= e^\beta d\vartheta,\;\;\; e^\varphi= 
e^\beta\sin\vartheta d\varphi,\nonumber\\
& & dx^a= e^{-\alpha} e^a,\;\;\; d\vartheta = e^{-\beta}e^\vartheta,\;\;\; 
d\varphi=\frac{e^{-\beta}}{\sin\vartheta} e^\varphi
\label{4}
\eea where $\beta \equiv f + \ln r_c.$ (Lower latin indeces from the
beginning of the alphabet indicate coordinates associted to the four
dimensional Minkowski spacetime with metric $\eta_{ab}$.)
Substituting into Eq.~(\ref{3}) we obtain 
\beq
\frac{G_3}{\kappa}=\epsilon\Big[b\frac{\cos\vartheta}{\sin\vartheta}
 e^{-2\alpha -\beta}  f_{ab}  e^\varphi \wedge e^a \wedge e^b - b
e^{-\alpha-2\beta} a_a e^\vartheta\wedge e^\varphi\wedge
e^{a}+\epsilon e^{-3\alpha}a_a f_{cb} e^{a}\wedge e^{c} \wedge e^{b}
\Big] \, ,
\label{5}
\eeq
where $f_{ab} = \partial_a a_b - \partial_b a_a$.
Because the three terms are orthogonal to each 
other straightforward calculation leads to
\beq
G^2_3= \kappa^2 \epsilon^2 (b^2\,\cot^2\vartheta \, e^{-4\alpha -2\beta} f_{ab}^2 + b^2 
e^{-2\alpha-4\beta} a_a^2)+ {\cal O}(\epsilon^4) \,.
\eeq
Then, the 5th term in Eq.~(\ref{action}) can be written as
\begin{eqnarray}
S_{G_3}  & = &- \frac{1}{2G_6}\, \int d^4 x \frac{G_6}{6}e^{4\alpha+2\beta}\sqrt{\eta_4}
e^{-2\phi} 
\int d\vartheta d\varphi \sin\vartheta \left[ \Big( \kappa^2 \epsilon^2 b^2\,\,\cot^2\vartheta  
e^{-4\alpha-2\beta} \Big) f_{ab}^2 \right. \nonumber \\
& + & \left. \Big( \kappa^2 \epsilon^2  b^2
e^{-2\alpha-4\beta}  \Big) a_a^2  \right] \, ,
\label{7}
\end{eqnarray}
whereas the contribution from the 4th term
in Eq.~(\ref{action}) can be computed from
Eq.~(\ref{Fdos}) yielding 
\bea
S_{F_2} &=&- \frac{1}{2G_6}\, \int d^4x \sqrt{\eta_4} 2 \pi e^{2\beta-\phi} G_6 \epsilon^2 
f_{ab}^2 \nonumber \\
&=&-\int d^4x \sqrt{\eta_4}  \pi e^{2f-\phi}r_c^2 \epsilon^2 f_{ab}^2 \,\, .
\eea
Thus,
\begin{equation}
S_{G_3} + S_{F_2} = - \int d^4x \left[\frac{1}{4\, g^2} f_{ab}^2 + \frac{m^2}{2}\, a_a^2 \right] \,,
\label{choco}
\end{equation}
where the four dimensional effective coupling  and the effective mass 
are of the form
\begin{equation}
\frac{1}{g^2}= 4\,  \epsilon^2 \sqrt{\eta_4} \left[
 \pi e^{2f - \phi} r_c^2 +
\frac{1}{12} \kappa^2  b^2 e^{-2\phi} \int d\vartheta d\varphi 
\sin\vartheta \cot^2\vartheta \right] \to \infty 
\end{equation}
and
\begin{equation}
m^2 = \frac{2}{3}\pi \k^2 b^2 \epsilon^2 e^{2\alpha-2\beta-2\phi} \, .
\label{9}
\end{equation}
For the moment we let $\int d\vartheta d\varphi 
\sin\vartheta \cot^2\vartheta = N$, where eventually we set $N \to \infty.$
Now to make quantum particle
identification and coupling, we carry out the transformation 
$a_a \to g \hat a_a$~\cite{note2}. This implies that the second term in the right hand side of Eq.~(\ref{choco}) vanishes, yielding
\begin{equation}
 f_{ab}  = \partial_a (g \hat a_b) - \partial_b (g \hat a_a) 
  =  \partial_a g \, \hat a_b - \partial_b g\, \hat a_a + g \, 
  \partial_a   \hat a_b - g\, \partial_b \hat a_a 
  =  g \hat f_{ab} + \hat a \wedge dg
\end{equation}
and consequently to leading order in $N$
\begin{equation}
\frac{1}{g^2} \, f_{ab}^2 = \frac{1}{g^2} [g^2 \hat f_{ab}^2 + (\hat a 
\wedge dg)^2 + 2\,g \, \hat a_b \,\, \hat f^{ab}\, \partial_a g] \,\, .
\end{equation}
If the coupling depends only  on the time variable, 
\begin{equation}
\frac{1}{g^2} \, f_{ab}^2  \to  \hat f_{ab}^2 + 
\left(\frac{\dot g}{g}\right)^2 \, \hat a_a^2  + 2 \, \frac{\dot g}{g} \,\, 
\hat a_i \,\, \hat f^{ti}
\end{equation}
where $\dot g = \partial_t g$ and lower latin indices from the middle
of the alphabet refer to the brane space-like dimensions.  If we
choose a time-like gauge in which $a_t = 0,$ then the term $(\dot
g/g)\, \hat a_i \, \hat f^{ti}$ can be written as $(1/2) (\dot g/g)
(d/dt) (\hat a_i)^2,$ which after an integration by parts gives $-
(1/2) [(d/dt) (\dot g/g)] \hat a_i^2$; with $g \sim e^{-\phi},$ the factor
in square brackets becomes $- \ddot \phi.$ Since $\phi =\sqrt{G_4} (X + Y),$
the rapidly varying $\ddot X$ will average to zero, and one is left
just with the very small $\ddot Y$, which is of order Hubble square. 
For the term $(\dot g/g)^2 (a_i)^2,$ the term $(\dot X)^2$ also averages to 
order Hubble square, implying that the induced mass term is of horizon size. 
These ``paraphotons''
carry new relativistic degrees of freedom, which could in turn modify
the Hubble expansion rate during Big Bang nucleosynthesis (BBN). Note,
however, that these extremely light gauge bosons are thought to be
created through inflaton decay and their interactions are only
relevant at Planck-type energies. Since the quantum gravity era, all
the paraphotons have been redshifting down without being subject to
reheating, and consequently at BBN they only count for a fraction of an
extra neutrino species in agreement with observations.


\begin{thebibliography}{99}


\bibitem{Riess:1998cb}
  A.~G.~Riess {\it et al.}  [Supernova Search Team Collaboration],
  Astron.\ J.\  {\bf 116}, 1009 (1998)
  [arXiv:astro-ph/9805201];
  S.~Perlmutter {\it et al.}  [Supernova Cosmology Project Collaboration],
  Astrophys.\ J.\  {\bf 517}, 565 (1999)
  [arXiv:astro-ph/9812133];
  N.~A.~Bahcall, J.~P.~Ostriker, S.~Perlmutter and P.~J.~Steinhardt,
  Science {\bf 284}, 1481 (1999)
  [arXiv:astro-ph/9906463].


\bibitem{Weinberg:dv}
S.~Weinberg,
Phys.\ Rev.\ Lett.\  {\bf 59}, 2607 (1987).


\bibitem{Bousso:2000xa}
  R.~Bousso and J.~Polchinski,
  JHEP {\bf 0006}, 006 (2000)
  [arXiv:hep-th/0004134];
L.~Susskind
arXiv:hep-th/0302219;
  M.~R.~Douglas,
  JHEP {\bf 0305}, 046 (2003)
  [arXiv:hep-th/0303194];
  N.~Arkani-Hamed and S.~Dimopoulos,
  JHEP {\bf 0506}, 073 (2005)
  [arXiv:hep-th/0405159];
  M.~R.~Douglas and S.~Kachru,
  arXiv:hep-th/0610102.



\bibitem{Maldacena:2000mw}
  J.~M.~Maldacena and C.~Nunez,
  Int.\ J.\ Mod.\ Phys.\  A {\bf 16}, 822 (2001)
  [arXiv:hep-th/0007018];
  G.~W.~Gibbons, ``Aspects of Supergravity Theories,'' lectures given at
  GIFT Seminar on Theoretical
  Physics, San Feliu de Guixols, Spain, 1984.
  Print-85-0061 (CAMBRIDGE), published in GIFT Seminar 1984:0123.




\bibitem{Gibbons:2001wy}
  G.~W.~Gibbons and C.~M.~Hull,
  arXiv:hep-th/0111072.

\bibitem{Townsend:2003fx}
  P.~K.~Townsend and M.~N.~R.~Wohlfarth,
  Phys.\ Rev.\ Lett.\  {\bf 91}, 061302 (2003)
  [arXiv:hep-th/0303097].
  See also,
  N.~Ohta,
  Phys.\ Rev.\ Lett.\  {\bf 91}, 061303 (2003)
  [arXiv:hep-th/0303238].



\bibitem{Giddings:2001yu}
  S.~B.~Giddings, S.~Kachru and J.~Polchinski,
  Phys.\ Rev.\  D {\bf 66}, 106006 (2002)
  [arXiv:hep-th/0105097].

%


\bibitem{Kachru:2003aw}
  S.~Kachru, R.~Kallosh, A.~Linde and S.~P.~Trivedi,
  Phys.\ Rev.\  D {\bf 68}, 046005 (2003)
  [arXiv:hep-th/0301240].





\bibitem{Salam:1984cj}
  A.~Salam and E.~Sezgin,
  Phys.\ Lett.\ B {\bf 147}, 47 (1984).



\bibitem{Cvetic:2003xr}
  M.~Cvetic, G.~W.~Gibbons and C.~N.~Pope,
  Nucl.\ Phys.\  B {\bf 677}, 164 (2004)
  [arXiv:hep-th/0308026].



\bibitem{Halliwell:1986bs} See e.g.,
  J.~J.~Halliwell,
  Nucl.\ Phys.\  B {\bf 286}, 729 (1987);
  Y.~Aghababaie, C.~P.~Burgess, S.~L.~Parameswaran and F.~Quevedo,
  JHEP {\bf 0303}, 032 (2003)
  [arXiv:hep-th/0212091];
  Y.~Aghababaie, C.~P.~Burgess, S.~L.~Parameswaran and F.~Quevedo,
  Nucl.\ Phys.\  B {\bf 680}, 389 (2004)
  [arXiv:hep-th/0304256];
  G.~W.~Gibbons, R.~Guven and C.~N.~Pope,
  Phys.\ Lett.\  B {\bf 595}, 498 (2004)
  [arXiv:hep-th/0307238];
  Y.~Aghababaie {\it et al.},
  JHEP {\bf 0309}, 037 (2003)
  [arXiv:hep-th/0308064].


\bibitem{Olive:1999ij}
  K.~A.~Olive, G.~Steigman and T.~P.~Walker,
  Phys.\ Rept.\  {\bf 333}, 389 (2000)
  [arXiv:astro-ph/9905320];
  R.~Bean, S.~H.~Hansen and A.~Melchiorri,
  Nucl.\ Phys.\ Proc.\ Suppl.\  {\bf 110}, 167 (2002)
  [arXiv:astro-ph/0201127].



\bibitem{Tegmark:2003ud}
  M.~Tegmark {\it et al.}  [SDSS Collaboration],
  Phys.\ Rev.\  D {\bf 69}, 103501 (2004)
  [arXiv:astro-ph/0310723].



\bibitem{Spergel:2006hy}
  D.~N.~Spergel {\it et al.}  [WMAP Collaboration],
  arXiv:astro-ph/0603449.




\bibitem{Halliwell:1986ja}
  J.~J.~Halliwell,
  Phys.\ Lett.\  B {\bf 185}, 341 (1987);
  B.~Ratra and P.~J.~E.~Peebles,
  Phys.\ Rev.\  D {\bf 37}, 3406 (1988);
  P.~G.~Ferreira and M.~Joyce,
  Phys.\ Rev.\ Lett.\  {\bf 79}, 4740 (1997)
  [arXiv:astro-ph/9707286];
  P.~G.~Ferreira and M.~Joyce,
  Phys.\ Rev.\  D {\bf 58}, 023503 (1998)
  [arXiv:astro-ph/9711102];
  E.~J.~Copeland, A.~R.~Liddle and D.~Wands,
  Phys.\ Rev.\  D {\bf 57}, 4686 (1998)
  [arXiv:gr-qc/9711068].



\bibitem{Comelli:2003cv}
  D.~Comelli, M.~Pietroni and A.~Riotto,
  Phys.\ Lett.\  B {\bf 571}, 115 (2003)
  [arXiv:hep-ph/0302080];
  U.~Franca and R.~Rosenfeld,
  Phys.\ Rev.\  D {\bf 69}, 063517 (2004)
  [arXiv:astro-ph/0308149].



\bibitem{Wald:1984rg}
  R.~M.~Wald,
  ``General Relativity,''
(University of Chicago Press, Chicago, 1984).



\bibitem{Vinet:2005dg} A similar expression was derived by
  J.~Vinet and J.~M.~Cline,
  Phys.\ Rev.\  D {\bf 71}, 064011 (2005)
  [arXiv:hep-th/0501098].





\bibitem{Preskill:1982cy}
  J.~Preskill, M.~B.~Wise and F.~Wilczek,
  Phys.\ Lett.\  B {\bf 120}, 127 (1983).







\bibitem{Hoffman:2003ru}
  M.~B.~Hoffman,
  arXiv:astro-ph/0307350.


\bibitem{note} This assumption will be justified {\em a posteriori}
  when we find that $\rho_X \ll \rho_{\rm mat}.$

\bibitem{copeland} E. J. Copeland, A. R. Liddle and D. Wands, {\it op. cit.} in Ref.~\cite{Halliwell:1986ja}.


\bibitem{Anchordoqui:2003ij}
  L.~Anchordoqui and H.~Goldberg,
  Phys.\ Rev.\  D {\bf 68}, 083513 (2003)
  [arXiv:hep-ph/0306084].

\bibitem{LopesFranca:2002ek}
  U.~J.~Lopes Franca and R.~Rosenfeld,
  JHEP {\bf 0210}, 015 (2002)
  [arXiv:astro-ph/0206194].


\bibitem{Steinhardt:1999nw}
  P.~J.~Steinhardt, L.~M.~Wang and I.~Zlatev,
  Phys.\ Rev.\  D {\bf 59}, 123504 (1999)
  [arXiv:astro-ph/9812313].

\bibitem{Carroll:1998zi}
  S.~M.~Carroll,
  Phys.\ Rev.\ Lett.\  {\bf 81}, 3067 (1998)
  [arXiv:astro-ph/9806099].



\bibitem{Aghababaie:2003wz}
  Y.~Aghababaie, C.~P.~Burgess, S.~L.~Parameswaran and F.~Quevedo,
  {\it op. cit.} in Ref.~\cite{Halliwell:1986ja}.

\bibitem{xcdm}
    This comes from a behavior $Y\simeq -\sqrt{2}u$
(compatible with the equations of motion), when
    combined with the $e^{-3u}$ in Eq.~(\ref{rowx}).


  \bibitem{cstring} Before proceeding, we remind the reader that the
    requirements for preserving a fraction of SUSY in spherical
    compactifications to four dimensions imply $b^2 \xi = 1$,
    corresponding to the winding number $n = \pm 1$ for the monopole
    configuration. In terms of the Bohm-Aharonov argument on phases,
    this is consistent with usual requirement of quantization of the
    monopole. The SUSY breaking has associated a non-quantized flux of
    the field supporting the two sphere. In other words, if we perform
    a  Bohm-Aharonov-like interference experiment, some phase change
    will be detected by a $U(1)$ charged  particle that circulates
    around the associated Dirac string. The quantization of fluxes
    implied the unobservability of such a phase, and so in our
    cosmological set up, the parallel transport of a fermion will be
    slightly  path dependent. One possibility is that the non-compact
    $\rho$ coordinate (in the uplift to ten dimensions, see Sec. III)
    is the direction in which the Dirac string exists. Then the cutoff
    necessary on the physics at large  $\rho$  will introduce a slight
    (time-dependent) perturbation on the flux quantization condition.
    We are engaged at present in exploring possibilities along this
    line.


\bibitem{note2} This is because the definition of the propagator with
  proper residue for correct Feyman rules in perturbation theory, and
  therefore also the couplings, needs to be consistent with the form
  of the Hamiltonian $= \sum_k \omega(k) a_k^\dagger a_k,$ with $[a,\,
  a^\dagger] =1$. This in turn implies that the kinetic term in the
  Lagrangian has the canonical form, $(1/4) \hat f_{ab}^2,$ with the usual
  expansion of the vector field $a_a.$




\end{thebibliography}
\end{document}